\documentclass[11pt,twoside,onecolumn]{article}
\usepackage[]{latexsym}
\usepackage{epsfig}
\usepackage{amsmath,amssymb}
\DeclareMathOperator\arctanh{arctanh}

\DeclareMathOperator\arccoth{arccoth}
\newcommand{\be}{\begin{eqnarray}}
\newcommand{\ee}{\end{eqnarray}}
\newcommand{\ud}{\mathrm{d}}

\newcommand{\gn}{G_{\rm N}}
\newcommand{\lp}{\ell_{\rm p}}
\newcommand{\mpl}{m_{\rm p}}

\newcommand{\expec}[1]{\mbox{$\langle #1\rangle$}}

\title{\bf Lorentz invariance without trans-Planckian physics?}
\author{Roberto~Casadio$^{a,b}$\thanks{roberto.casadio@bo.infn.it}
\\
\null
\\
$^a${\em Dipartimento di Fisica e Astronomia, Universit\`a di Bologna}
\\
{\em via Irnerio~46, 40126 Bologna, Italy}
\\
\\
$^c${\em Istituto Nazionale di Fisica Nucleare, Sezione di Bologna}
\\
{\em via Irnerio~46, 40126 Bologna, Italy}
}
\begin{document}
\maketitle
\begin{abstract}
We explore the possibility that, in a quantum field theory with Planck scale
cutoff $\Lambda\simeq\mpl$, observable quantities for low-energy processes
respect the Lorentz symmetry. 
In particular, we compute the one-loop radiative correction $\Pi$ to the
self-energy of a scalar field with $\lambda\,\phi^4$ interaction, using a modified
(non-invariant) propagator which vanishes in the trans-Planckian
regime, as expected in the ``classicalization'' scenario.
We then show that, by imposing the result does not depend on $\Lambda$
(in the limit $\Lambda\to\mpl$), an explicit (albeit not unique)
expression for $\Pi$ can be derived, which is similar to the one
simply obtained with the standard Feynman propagator and a cutoff
$\Lambda=\mpl$.
\end{abstract}
%
%
%
\maketitle
\section{Introduction}
\label{secIntro}
\setcounter{equation}{0}
It is usually believed that quantum gravitational effects should become
relevant at energy scales of the order of the Planck mass,
$\mpl\simeq 10^{16}\,$TeV, or higher.
This conclusion is easily reached by considering that the Einstein-Hilbert
action is proportional to the Newton constant $\gn=\lp/\mpl$~\footnote{We
shall always use units with $c=1$ and $\hbar=\lp\,\mpl$.},
and gravitational perturbations on a given background therefore
couple to matter sources with a strength proportional to
$\lp/\mpl\simeq\mpl^{-2}$.
The strength of the matter-gravity coupling can also be seen directly
in the semiclassical Einstein field equations,  
\be
R_{\mu\nu}-\frac{1}{2}\,R\,g_{\mu\nu}
=
8\,\pi\,\gn\,\expec{\hat T_{\mu\nu}}
\ ,
\label{Einstein}
\ee
where the expectation value $\expec{\hat T_{\mu\nu}}$ of the energy-momentum
(operator) tensor on a given quantum state of matter has replaced its classical
counterpart $T_{\mu\nu}$.
\par
A clear exception is given by purely classical vacuum solutions of
Eq.~\eqref{Einstein}, for which $\expec{\hat T_{\mu\nu}}\simeq T_{\mu\nu}=0$,
so that $\gn$ apparently drops from the calculation.
In fact, $\gn$ can re-enter as part of an integration constant proportional to
the mass $m$ of a spin-less point-like source, and turns it into a length, 
namely the Schwarzschild radius
\be
R_M
=
2\,\gn\,m
\equiv
2\,M
\ .
\ee
On the other hand, for such a particle, quantum mechanics introduces
an uncertainty in spatial localisation, typically of the order of the
Compton (de~Broglie) length,
\be
\lambda_M
\simeq
\frac{\lp\,\mpl}{m}
=
\frac{\lp^2}{M}
\ .
\label{lambdaM}
\ee
Given that quantum physics is a more refined description of reality than classical physics,
the clash of the two lengths, $R_m$ and $\lambda_m$, implies that the former
only makes sense provided it is significantly larger than the latter,
\be
R_M\gtrsim \lambda_M
\quad
\Rightarrow
\quad
m
\gtrsim
\mpl
\ ,
\label{clM}
\ee
or $M\gtrsim\lp$.
Note that this argument employs the flat space Compton length~\eqref{lambdaM},
and it is likely that the particle's self-gravity will affect it.
However, it is still reasonable to assume the condition~\eqref{clM} holds as a rough,
order of magnitude, estimate. 
In fact, one can alternatively consider the ``mean energy density'' inside the
Schwarzschild radius,
\be
{\mathcal E}_H
\simeq
\frac{m}{R_H^3}
=
\frac{\mpl^3}{\lp^3\,m^2}
\ ,
\ee
and require that it does not exceed the Planck scale,
\be
{\mathcal E}_H
\lesssim
\lp^{-3}\,\mpl
\ ,
\ee
which again leads to Eq.~\eqref{clM}.
\par
Overall, the above-mentioned consideration that quantum gravity effects
become relevant for $m$ of order $\mpl$ or higher now appears questionable,
since the condition~\eqref{clM} implies that such a system should be fairly well
described in classical terms.
This is indeed at the core of the recent ideas of UV self-completeness of gravity
and ``classicalization''~\cite{dvali}, as well as it had previously inspired
Generalized Uncertainty Principles (GUPs)~\cite{gup}.
More or less implicitly, such scenarios require the existence of a preferred (inertial)
reference frame in which the components of four-momenta reach Planck size,
thus breaking Lorentz covariance at face value~\footnote{For a recent approach that
employs a Lorentz covariant cutoff, see~\cite{kempf} and References therein.}.
Our main aim in this work will be precisely to investigate the possibility that
Lorentz symmetry at low-energy and ``classicalization'' --  or, more precisely,
a total suppression of trans-Planckian quantum modes, can be effectively
reconciled.
\section{Gravitational renormalization}
\label{secGrRen}
\setcounter{equation}{0}
There are many reasons to indulge in the possibility that the gravitational interaction
causes Quantum Field Theory (QFT) propagators to depart from their flat-space
expressions at high energy.
Whatever the reason, we then need an explicit implementation in order to compute
physical predictions. 
Classicalization induced by black hole formation and the GUPs are proposals we
already mentioned above.
Alternatively, a set of ``diagrammatic rules'' was presented
in Ref.~\cite{casadio} to effectively (and non-perturbatively) include (self-)gravity
in the standard perturbative QFT of matter and other interactions.
Since such rules will not be explicitly needed here, we will just recall
the basic idea:
in Feynman diagrams, each flat-space Feynman propagator of momentum $p$,
should be replaced by the propagator in the curved space-time sourced by all
the other (real or virtual) particles (say, with total momentum $q$) in the same diagram,
to wit
\be
G(p)
\rightarrow
G_q(p;\mpl)
\ ,
\label{Gqp}
\ee
where we also allowed for an explicit dependence on the
Planck scale $\mpl$, as a reminder that (self-)gravity is to be included.
These propagators could, in principle, be obtained perturbatively,
by summing over infinitely many graviton exchange diagrams or,
non-perturbatively -- but perhaps, equivalently, by solving the
semiclassical equations~\eqref{Einstein}, although this task is likely
unattainable without some other approximation.
For example, in Ref.~\cite{casadio}, a modified scalar propagator was
derived, under the working assumption that the Schwarzschild metric can
be approximated by a conformally flat metric for (short-lived) virtual processes. 
The one-loop correction to the four-point correlation function
for the scalar field with $\lambda\,\phi^4$ interaction
was then shown to contain no Ultra-Violet (UV) divergences. 
\par
In the following, we shall assume that classicalization works, with no quantum
degree of freedom propagating above the Planckian scale, and just
focus on the requirements the propagator must consequently satisfy to build a
theory consistent with low-energy Lorentz symmetry.
To this purpose, we shall not (totally) specify the modified propagator in
Eq.~\eqref{Gqp}, but assume that when any component $k^\mu$ of the internal
momenta reaches the Planck scale [so that condition~\eqref{clM} is roughly satisfied],
a classical configuration forms, whose contribution as a radiative correction
is negligible.
This assumption can be effectively formalised by introducing a UV cutoff $\Lambda$
in the integrals over components of the virtual momenta at the Planck scale,
say $\Lambda\simeq\mpl$.
This rule also seems to require a preferred reference frame.
For example, one may consider the rest frame of the (virtual) forming
black hole, in which the spatial components of its four-momentum 
vanish, that is $k^\mu=(E,0,0,0)$, and apply a continuous change of frame
while performing the integration over virtual momenta~\footnote{This possibility
is currently being investigated, but appears technically very involved.}.
We shall here opt for a simpler picture, we are now going
to illustrate with an example.
\section{Gravitationally renormalised self-energy}
\label{secSelf}
\setcounter{equation}{0}
\begin{figure}[t]
\centering
$k^\mu=(E,p^i)$
\\
$P^\mu$
\includegraphics[width=5cm]{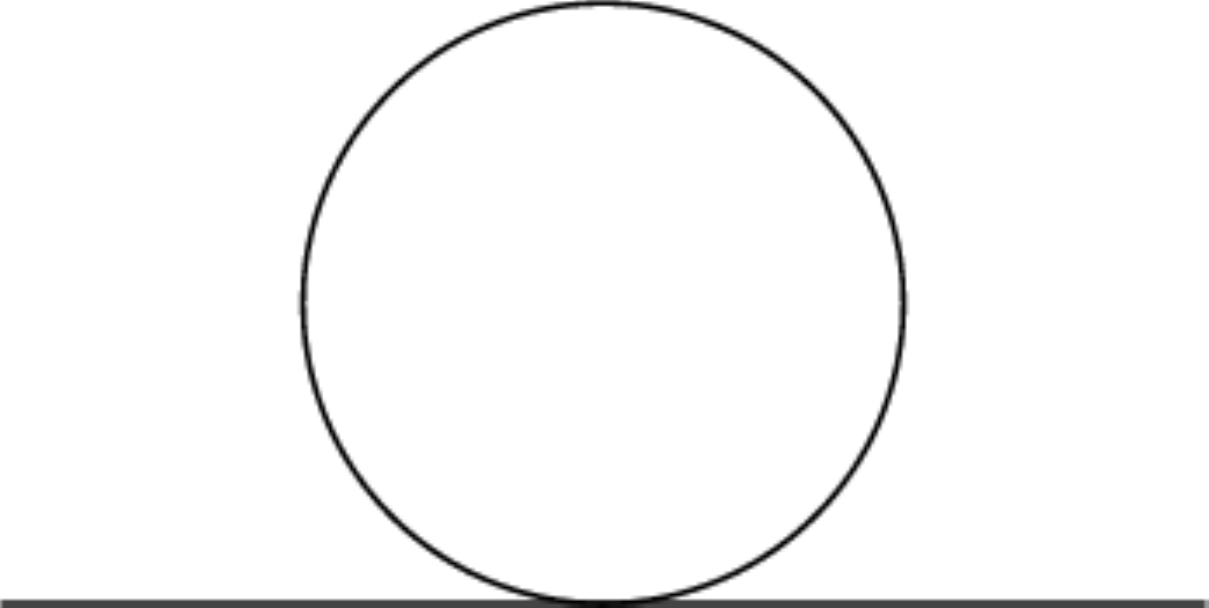}
$P^\mu$
\caption{One-loop correction to the mass from $\lambda\,\phi^4$.
\label{mass}}
\end{figure}
We wish to test the above rule on the self-energy
of a scalar field with $\lambda\,\phi^4$ interaction.
We shall first assume, for the sake of simplicity, that there exists a global
inertial frame in which the cutoff is isotropic, and then estimate
the possible corrections induced by local fluctuations of the cutoff itself.
\subsection{Global isotropic cutoff}
We shall here assume there exists a global inertial reference frame
where the momentum cutoff is given by the isotropic four-vector
\be
\Lambda^\alpha=(\Lambda,\Lambda,\Lambda,\Lambda)
\ .
\label{isoL}
\ee
The one-loop contribution is therefore represented by the tadpole diagram
(see Fig.~\ref{mass}), and reads
\be
\Pi_{\Lambda}(P^2)
=
\lambda
\left(
\prod_{\mu=0}^3
\int\limits_{-\Lambda}^{+\Lambda}
\ud k^\mu
\right)
G_P(k;\mpl)
\ ,
\ee
where $P^\mu$ is the (on-shell) four-momentum of the scalar particle,
with $P^2\equiv P_\mu\,P^\mu=m_0^2$, and $G_P(k;\mpl)$ is the modified
propagator from Eq.~\eqref{Gqp}.
In a different inertial frame, the cutoff four-vector will be
$\bar \Lambda^\alpha=L^\alpha_{\ \beta}\,\Lambda^\beta$
(where $L$ is a Lorentz matrix), and we must likewise have
\be
\Pi_{\bar \Lambda}(P^2)
=
\lambda
\left(
\prod_{\mu=0}^3
\int\limits_{-\bar \Lambda^\mu}^{+\bar \Lambda^\mu}
\ud k^\mu
\right)
G_P(k;\mpl)
\ ,
\ee
where $\bar P^\alpha=L^\alpha_{\ \beta}\,P^\beta$ is the boosted external
momentum, again with $\bar P^2=m_0^2$.
If the result has to be invariant under (small) changes of the cutoff,
$\Pi_{\Lambda}(P^2)=\Pi_{\bar \Lambda}(P^2)$ for
$\Lambda\simeq\bar\Lambda\sim\mpl$,
at least when the components $|P^\mu|\ll\mpl$, we must then have
\be
\left.
\frac{\partial \Pi_{\Lambda}(P^2)}{\partial \Lambda}
\right|_{\Lambda=\mpl}
=
0
\ ,
\label{dPidL}
\ee
which can be more explicitly written as
\be
\sum_{\mu=0}^3
\left[
\left(
\prod_{\nu\not=\mu}
\int\limits_{-\mpl}^{+\mpl}
\ud k^\nu
\right)
G_P(k;\mpl)
\right]_{k^\mu=\pm\mpl}
\!\!\!\!\!\!\!\!
=0
\ .
\label{cMpl}
\ee
Clearly, Eq.~\eqref{cMpl} would hold if the modified propagator $G_P(k;\mpl)$
vanished when the components $k^\mu=\pm\mpl$, and does not hold with
the usual Feynman propagator~\footnote{The proposed propagator in
Ref.~\cite{casadio} looks marginally better, due to the
suppression weight $\rho_\Lambda(k)$, as does the exponentially suppressed
propagator obtained from non-commutativity in Ref.~\cite{spallucci}.
The latter has also the clear advantage of being explicitly covariant in form,
albeit in the Euclidean formulation (after a Wick rotation that maps time to imaginary
values).}.
Further, since we are interested in the low-energy regime for the
external particles, all the components $|P^\mu|\ll\mpl$, and in Eq.~\eqref{cMpl}
we can approximate
\be
G_P(k;\mpl)
\simeq
G_{P=0}(k;\mpl)
\equiv
G(k^\mu;\mpl)
\ ,
\ee
where Greek indices run from $0$ to $3$
(Latin indices $i=1,2,3$ and $a=1,2$), and obtain
\be
\sum_{\mu=0}^3
\left[
\left(
\prod_{\nu\not=\mu}
\int\limits_{-\mpl}^{+\mpl}
\ud k^\nu
\right)
G(k^\alpha;\mpl)
\right]_{k^\mu=\pm\mpl}
\!\!\!\!\!\!\!\!
=0
\ .
\ee
Since the ``preferred'' direction $P^\mu$ was dropped, we can now employ
homogeneity of Minkowski space-time in order to write the above as
\be
0
&\!\!\simeq\!\!&
2\,\int\limits_{-\mpl}^{+\mpl}
\!\!
\ud k^1\,\ud k^2\,\ud k^3
\left.
G(k^\mu;\mpl)
\right|_{k^0=\mpl}
+6
\int\limits_{-\mpl}^{+\mpl}
\!\!
\ud k^0\,\ud k^1\,\ud k^2
\left.
G(k^\mu;\mpl)
\right|_{k^3=\mpl}
\nonumber
\\
&\!\!=\!\!&
8\,\pi
\int_0^{\mpl}
\!\!
p^2\,\ud p
\,
G(\mpl,p^i;\mpl)
+
12\,\pi
\int_0^{\mpl}
\!\!
\ud E
\int_0^{\mpl}
\!\!
q\,\ud q
\,
G(E,q^a,\mpl;\mpl)
\nonumber
\\
&\!\!\equiv\!\!&
I_1+I_2
\ .
\label{cc}
\ee
We further assume 
\be
G(k^\mu;\mpl)
=
\frac{g(E,p;\mpl)}{E^2-p^2-m_0^2}
\ ,
\label{prop}
\ee
where the function $g(E\lesssim\mpl,p\lesssim\mpl;\mpl)\simeq 1$,
in order to recover the standard Feynman propagator at low momenta.
The only non-vanishing contribution to the right hand side of
Eq.~\eqref{cc} then comes from values of the integrands around $\Lambda\sim \mpl$,
namely $E\simeq p\simeq\mpl$.
In fact,
\be
I_1(g=1)
&\!\!=\!\!&
8\,\pi
\int_0^{\Lambda}
\!\!
\frac{p^2\,\ud p}{\mpl^2-p^2-m_0^2}
\nonumber
\\
&\!\!\simeq\!\!&
8\,\pi\left[
\mpl\,\arctanh\left(\frac{\Lambda}{\mpl}\right)
-\Lambda
\right]
\nonumber
\\
&\!\!\simeq\!\!&
4\,\pi\,\mpl\,\ln\!\left(\frac{\mpl}{\mpl-\Lambda}\right)
+\mathcal{O}(\mpl-\Lambda)
\ ,
\label{lhs11}
\ee
for $\Lambda\to\mpl^-$, and neglecting the bare mass $m_0\ll\mpl$.
Likewise,
\be
I_2(g=1)
&\!\!=\!\!&
12\,\pi
\int_0^{\Lambda}
\!\!
\ud E
\int_0^{\Lambda}
\!\!
\frac{q\,\ud q}{E^2-q^2-\mpl^2-m_0^2}
\nonumber
\\
&\!\!\simeq\!\!&
12\,\pi\,\mpl\,\arctanh\left(\frac{\Lambda}{\mpl}\right)
-12\,\pi\,\sqrt{\mpl^2+\Lambda^2}\,
\arctanh\left(\frac{\Lambda}{\sqrt{\mpl^2+\Lambda^2}}\right)
\nonumber
\\
&&
+12\,\pi\,\Lambda\,
\ln\left(\frac{\mpl^2-\Lambda^2}{\mpl^2}\right)
\nonumber
\\
&\!\!\simeq\!\!&
12\,\pi\,\mpl
\left[\ln(2)-\sqrt{2}\,\arccoth(\sqrt{2})\right]
+\mathcal{O}(\mpl-\Lambda)
\ .
\label{lhs21}
\ee
Since Eq.~\eqref{lhs11} diverges for $\Lambda\to\mpl$,
the function $g$ must be at least of order $(\mpl-\Lambda)$,
for $p\sim q\sim E\sim \Lambda$,
in order to cure the divergence and satisfy~\eqref{cc}.
\par
We shall therefore replace all the UV cutoffs at $\mpl$ with a general
$\mpl^2\gtrsim\Lambda^2\gg P^2$, assume
\be
g(E,p^i;\mpl)
&\!\!=\!\!&
\frac{1}{\alpha}
\left[
\frac{\mpl^2-p^2}{\mpl^2}
+
(\alpha-1)\,
\frac{\mpl^2-E^2}{\mpl^2}
\right]
\nonumber
\\
&\!\!=\!\!&
1+\frac{k^2-\alpha\,E^2}{\alpha\,\mpl^2}
+\mathcal{O}\left(\frac{k^3}{\mpl^3}\right)
\ ,
\label{g}
\ee
and take the limit $\Lambda\to\mpl^-$ at the end of the calculation.
From this ansatz, we obtain
\be
\frac{I_1}{4\,\pi}
&\!\!=\!\!&
\frac{m_0^2}{\alpha\,\mpl^2}\,\sqrt{\mpl^2-m_0^2}\,
\arctanh\!\left(\frac{\Lambda}{\sqrt{\mpl^2+m_0^2}}\right)
+\frac{\Lambda}{3\,\alpha\,\mpl^2}
\left(
\Lambda^2
-3\,m_0\right)
\nonumber
\\
&\!\!=\!\!&
\frac{\Lambda^3}{3\,\alpha\,\mpl^2}
+\mathcal{O}\left(\frac{m_0^2}{\mpl^2}\right)
\ ,
\label{lhs1}
\ee
and
\be
\frac{I_2}{\pi}
&\!\!=\!\!&
\frac{\Lambda}{\mpl^2}
\!
\left[
\frac{3-2\,\alpha}{\alpha}\,\Lambda^2
+(\Lambda^2-3\,\mpl^2)
\ln\!\left(\frac{\mpl^2-\Lambda^2}{\mpl^2}\right)
\!
\right]
\nonumber
\\
&&
+
\frac{2\,(\Lambda^2-2\,\mpl^2)}
{\mpl^2}\,\sqrt{\mpl^2+\Lambda^2}
\arctanh\!\left(\frac{\Lambda}{\sqrt{\mpl^2+\Lambda^2}}\right)
\nonumber
\\
&&
+
4\,\mpl
\arctanh\!\left(\frac{\Lambda}{\mpl}\right)
+\mathcal{O}\left(\frac{m_0^2}{\mpl^2}\right)
\ .
\label{lhs2}
\ee
Taking the limit $\Lambda\to\mpl^-$ and substituting $I_1$ and
$I_2$ into Eq.~\eqref{cc} therefore yields an equation for the parameter
$\alpha$, which can be easily solved, that is
\be
\alpha
=
\frac{13}
{6\left[
1+\sqrt{2}\,\arccoth(\sqrt{2})-2\,\ln 2\right]}
\simeq
2.5
\ ,
\ee
or
\be
g(E,p^i;\mpl)
&\!\!\simeq\!\!&
0.4
\left(
\frac{\mpl^2-E^2}{\mpl^2}
+
1.5\,\frac{\mpl^2-p^2}{\mpl^2}
\right)
\nonumber
\\
&\!\!\simeq\!\!&
1+0.4\,\frac{k^2-2.5\,E^2}{\mpl^2}
\ .
\ee
Note that the function $g$ does not only depend on the Lorentz scalar
$k^2=E^2-p^2$, but also on the energy $E$.
The propagator $G_P(k;\mpl)$ is therefore not a Lorentz scalar.
This seems a necessary price to pay in order to compensate for the
Lorentz dependence of the cutoff $\Lambda$, and perhaps not such 
a costly one, since the propagator is not an observable
{\em per se\/}~\footnote{Strictly speaking, the self-energy is hardly observable
either, however we chose this quantity as a reasonably simple toy case to
test our line of reasoning.}.
More specifically, the constant $\alpha\not= 0$ signals the departure
(of order $E^2/\mpl^2$) of $G_P(k;\mpl)$ from being a Lorentz scalar.
On the other end, the function $g$ is singular in the limit $\alpha\to 0$
for $\mpl$ finite, and the Lorentz violating correction does not depend
on $\alpha$ in the low-energy limit. 
One may therefore argue that fixing the UV scale will bring down 
necessary modifications to the low energy regime (a form of IR-UV mixing).
\par
With the condition~\eqref{cc} satisfied, we can finally estimate the
mass correction, namely
\be
\Pi_{\mpl}(m_0^2)
&\!\!=\!\!&
\lambda
\lim_{\Lambda\to\mpl}
\int_0^\Lambda
\ud^4 k\,
\frac{g(k^\mu;\mpl)}{k^2-m_0^2}
\nonumber
\\
&\!\!=\!\!&
2\,\frac{\pi}{3}
\left(\frac{2}{\alpha}-3\right)
\lambda\,\mpl^2
\left[1
+\mathcal{O}\left(\frac{m_0^2}{\mpl^2}\right)
\right]
\nonumber
\\
&\!\!\simeq\!\!&
-4.6\,\lambda\,\mpl^2
\left[1+\mathcal{O}\left(\frac{m_0^2}{\mpl^2}\right)\right]
\ ,
\label{Pi}
\ee
where the Planck mass $\mpl$ must here be viewed as a universal constant.
The result is therefore a (low-energy, $m_0\ll\mpl$) Lorentz scalar,
like we wanted.
Of course, one might argue that the chosen form of the weight function
$g$ in Eq.~\eqref{g} is hardly the unique solution for the constraint~\eqref{cc},
and the final expression~\eqref{Pi} remains consequently ambiguous.
However, if we compare with the result derived by using the
standard Feynman propagator ($g=1$),
\be
\Pi_\Lambda(m_0^2)
&\!\!=\!\!&
\lambda
\int_0^\Lambda
\frac{\ud^4 k}{k^2-m_0^2}
\nonumber
\\
&\!\!=\!\!&
-2\,\pi\,\lambda\,\Lambda^2
\left[1
+\mathcal{O}\left(\frac{m_0^2}{\Lambda^2}\right)
\right]
\ ,
\label{stPi}
\ee
and set $\Lambda=\mpl$, we see that we obtained a correction
of the same form.
The fact that our result~\eqref{Pi} closely resembles~\eqref{stPi}
is suggestive that, perhaps, any reasonably behaved modified propagator
$G_P(k;\mpl)$ which solves~\eqref{cc} would lead to the same kind
of mass correction.
Eq.~\eqref{Pi} also implies that $|\Pi_{\mpl}|\sim\mpl^2\gg m_0^2$, unless
$\lambda\sim\mpl^{-2}$, and one must still apply 
the usual subtraction at the renormalisation point in order to set the
mass $\mu^2\simeq m_0^2-\Pi$ to the ``observed value''. 
\subsection{Fluctuating cutoff}
One might question the existence of a global reference frame in which the
momentum cutoff takes the isotropic form in Eq.~\eqref{isoL}.
For example, there are models in which the space-time appears as a foam
(of virtual black holes) at the microscopic level~\footnote{The literature on this
subject is rather extensive (see, for instance, Refs.~\cite{foam}).},
and it is therefore reasonable to consider a stochastic dependence of the cutoff four-vector
on position and time.
\par
Previous results should then be corrected, for example, by taking an
``ensemble average'' over the stochastic distribution of cutoff around
the Planck mass.
This means that Eq.~\eqref{dPidL} should be replaced by
\be
\left\langle
\frac{\partial \Pi_{\Lambda}(P^2)}{\partial \Lambda}
\right\rangle
\equiv
\int
\ud m\,
F_{\mpl}(m)
\left.
\frac{\partial \Pi_{\Lambda}(P^2)}{\partial \Lambda}
\right|_{\Lambda=m}
=
0
\ ,
\label{<dPidL>}
\ee
where $F_{\mpl}$ is a distribution peaked around the Planck scale $m\sim\mpl$
that could be specified given a microscopic model of the space-time,
and after integration on the angular variables (to restore local isotropy).
It is then easy to see that the final result~\eqref{Pi} becomes,
at least to leading order in $m_0/\mpl$,
\be
\Pi_{\mpl}(m_0^2)
&\!\!\simeq\!\!&
-4.6\,\lambda
\left[1+\mathcal{O}\left(\frac{m_0^2}{\mpl^2}\right)\right]
\int
\ud m\,
F_{\mpl}(m)\,
m^2
\nonumber
\\
&\!\!\simeq\!\!&
-\beta\,\lambda\,\mpl^2
\left[1+\mathcal{O}\left(\frac{m_0^2}{\mpl^2}\right)\right]
\ ,
\label{<Pi>}
\ee
where $\beta$ is just a numerical coefficient (of order one) that depends on the details
of the stochastic distribution $F_{\mpl}$.
\par
To conclude, it is rather unlikely that the form of $F_{\mpl}$ is such that subtle cancellations
occur in Eq.~\eqref{dPidL}, so as to drastically change the final result, and we do not expect
any significant modifications from the (more realistic) picture of a space-time dependent cutoff.
\section{Final remarks}
\label{secFin}
\setcounter{equation}{0}
We have shown that, in the simple case of a (massive) scalar field,
the self-energy correction $\Pi$ can be computed in a QFT with a cutoff at the
Planck scale $\mpl$, and still obtain a Lorentz invariant result
by modifying the propagator:
the two non-invariances (of the cutoff and of the propagator) compensate
each other and give rise to a (low-energy) frame-independent $\Pi$.
Such quantity naturally depends on $\mpl$, which must be viewed
as a universal (frame-independent) constant.
Also, the correction differs just by numerical coefficients
from the $\Pi$ obtained from the usual Feynman propagator,
which suggests that, if modifications to the propagator can be
related to the scalar field self-gravitational interaction, the effect
of the latter should be mild in this context.
And that quantum gravitational effects might indeed have an almost
irrelevant phenomenological impact on Standard Model predictions
to all scales.  
\par
Of course, the above result is far from sufficient to prove that the 
question raised in the title of this letter can be answered positively.
For that purpose, one should generalise the above procedure and
require Lorentz invariance of all quantities we can observe in
particle physics (such as scattering cross-sections, etc.).
In order to achieve this, it will be necessary that the deforming weight
$g$ in the propagator~\eqref{prop} contains enough degrees
of freedom (or parameters, like $\alpha$ in the example
above) to satisfy the equivalent of Eq.~\eqref{cc}.
This should not be difficult to accommodate in the spirit of the GUPs
or of the rules of Ref.~\cite{casadio}, since in diagrams with $N$ internal lines,
each corresponding propagators should depend (at least) on the $N-1$
other virtual particles in the graph (and external real particles),
and one expects to have at least $N-1$ such parameters.
\par
Finally, but not less important, there remains to see if gauge invariances
and other symmetries of the Standard Model can be preserved as well,
after imposing the low-energy Lorentz invariance of observable quantities.
Addressing this crucial issues requires investigating more realistic gauge
QFTs, rather than toy model scalar fields, and, unless one can find
a systematic procedure, it will also involve a significant amount of work.  
\end{document}